\documentclass[sigconf,authorversion]{acmart}
\AtBeginDocument{%
  \providecommand\BibTeX{{%
    \normalfont B\kern-0.5em{\scshape i\kern-0.25em b}\kern-0.8em\TeX}}}

\copyrightyear{2022}
\acmYear{2022}
\setcopyright{rightsretained}
\acmConference[MuC '22]{Mensch und Computer 2022}{September 4--7, 2022}{Darmstadt, Germany}
\acmBooktitle{Mensch und Computer 2022 (MuC '22), September 4--7, 2022, Darmstadt, Germany}
\acmDOI{10.1145/3543758.3543766}
\acmISBN{978-1-4503-9690-5/22/09}

\usepackage{algorithm} 
\usepackage{algorithmic} 
\begin{document}

\title[The Gesture Authoring Space: Custom Tailored Hand Gestures]{The Gesture Authoring Space: Authoring Customised Hand Gestures for Grasping Virtual Objects in Immersive Virtual Environments}

\author{Alexander Schäfer}
\email{alexander.schaefer@dfki.de}
\orcid{1234-5678-9012}
\affiliation{%
  \institution{TU Kaiserslautern}
  \streetaddress{P.O. Box 1212}
  \city{Kaiserslautern}
  \state{Rheinland Pfalz}
  \country{Germany}
  \postcode{66953}
}

\author{Gerd Reis}
\affiliation{%
  \institution{German Research Center for Artificial Intelligence}
  \streetaddress{Trippstadterstraße 122}
  \city{Kaiserslautern}
  \country{Germany}}
\email{gerd.reis@dfki.de}

\author{Didier Stricker}
\affiliation{%
  \institution{German Research Center for Artificial Intelligence, TU Kaiserslautern}
  \city{Kaiserslautern}
  \country{Germany}
}

\renewcommand{\shortauthors}{Schäfer et al.}

\begin{abstract}
Natural user interfaces are on the rise. Manufacturers for Augmented, Virtual, and Mixed Reality head mounted displays are increasingly integrating new sensors into their consumer grade products, allowing gesture recognition without additional hardware. This offers new possibilities for bare handed interaction within virtual environments. This work proposes a hand gesture authoring tool for object specific grab gestures allowing virtual objects to be grabbed as in the real world. The presented solution uses template matching for gesture recognition and requires no technical knowledge to design and create custom tailored hand gestures. In a user study, the proposed approach is compared with the pinch gesture and the controller for grasping virtual objects. The different grasping techniques are compared in terms of accuracy, task completion time, usability, and naturalness. The study showed that gestures created with the proposed approach are perceived by users as a more natural input modality than the others.
\end{abstract}

\ccsdesc[500]{Human-centered computing~Gestural input}
\ccsdesc[300]{Human-centered computing~Virtual reality}
\ccsdesc[300]{Human-centered computing~Mixed / augmented reality}

\keywords{virtual reality, augmented reality, mixed reality, hand gestures, bare handed, gesture, virtual object grabbing, grasping, manipulation}

\begin{teaserfigure}
\centering
  \includegraphics[width=0.8\textwidth]{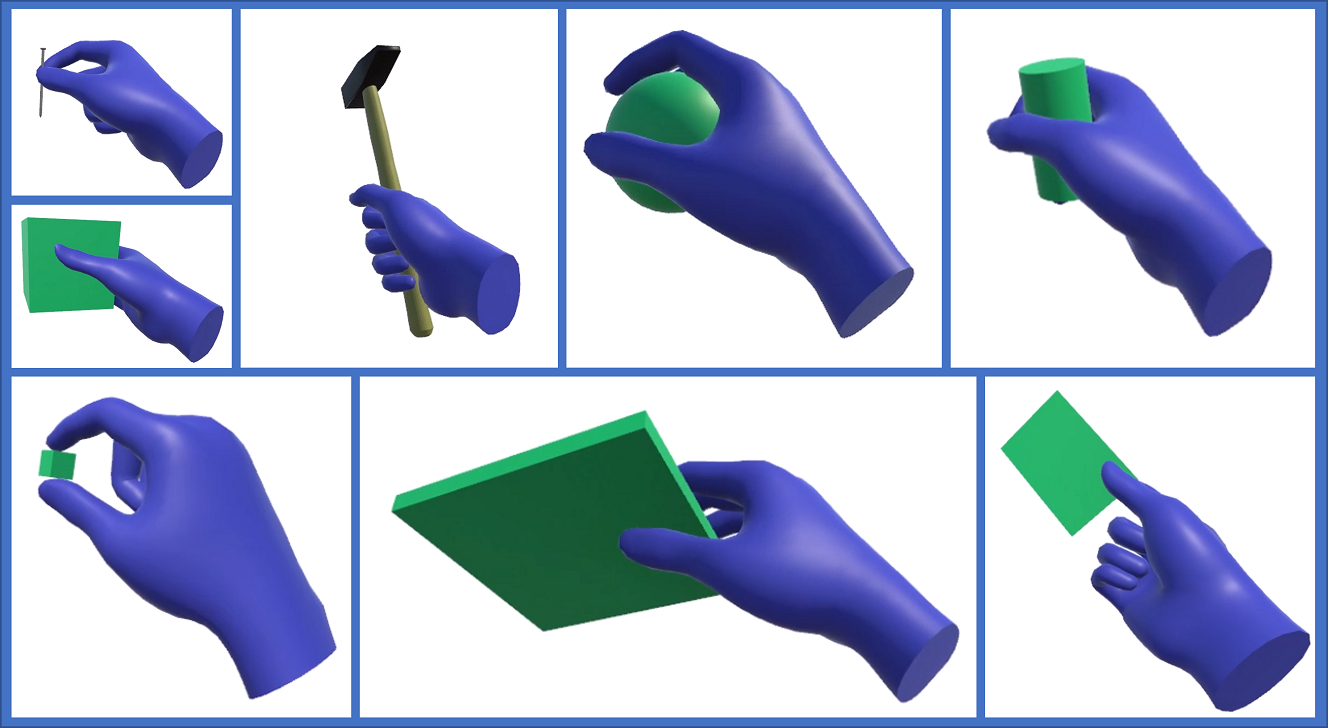}
  \caption{The proposed framework allows users to create their own custom tailored hand gestures for grasping virtual objects. The framework was evaluated with a user study and the grasping is compared to picking up with a controller and the pinch gesture.}
  \Description{}
  \label{fig:teaser}
\end{teaserfigure}

\maketitle

\section{Introduction}
\label{Chapter:GestureAuthoring:sec:Introduction}
Gesture recognition and hand-based interaction is becoming increasingly important for Augmented (AR), Virtual (VR), and Mixed Reality (MR) applications. Hand gestures have a wide range of application scenarios. For example, they are used in an AR experiment by \citet{iqbal2021exploring} to learn chemical reactions by utilizing the real hands. Hand gestures are also commonly used to learn and recognise sign language as done by \citet{shin2021american}. 
Virtual hands are also used as communication cues in a remote collaboration scenario using 360° images such as done by \citet{teo2018hand}.
Hand gestures are also used in VR medical training scenarios by \citet{khundam2021comparative} to pick up objects and interact with elements in the virtuel environment. \par
An essential interaction for AR, VR, and MR applications is virtual object manipulation. Surveys about applications with such immersive technologies show the importance of this interaction. In the work of \citet{10.1145/3533376} virtual object manipulation is ranked as the most common interaction type for remote collaboration scenarios. The most common way to pick up virtual objects is to press a button on a VR controller. With the recent rise of available hand tracking in affordable head mounted displays (HMDs) for AR, VR, and MR applications, researchers and practitioners are exploring  different ways to pick up objects. Without using controllers, the pinch gesture is the most common gesture to pick up objects. The software development kits for HMDs with hand tracking usually include a pinch gesture as the default gesture to pick up objects, as for example with the Hololens 2\footnote{https://docs.microsoft.com/windows/mixed-reality/mrtk-unity/}, Meta Quest\footnote{https://developer.oculus.com/}, or the Leap Motion\footnote{https://developer.leapmotion.com/, last accessed at 29.06.2022} to name the most common examples. An early example which used the pinch gesture to grasp virtual objects was introduced by \citet{fingARtips}. As the pinch gesture is easy to recognise and  can be reliably performed by users, it is the preferred method for vendors to showcase the capability of this technology. However, the pinch gesture is not optimal for many use cases. First, it is an unnatural gesture to pick up many objects because of their geometric properties. Second, the thumb and index finger need to be close to each other for a pinch gesture. This prevents some other gestures which require the thumb and index finger being close to each other from being recognised. \par 
This work aims to improve the current state of the art for picking up virtual objects with bare hands. Previous studies show that users, and especially lay people to this technology, often try to grab virtual objects as if they were picking up objects in the real world such as observed by \citet{kang_comparative_2020} and \citet{schafer2022comparing}. It is investigated how the current state of the art for picking up virtual objects with bare hands, the pinch gesture, can be replaced by more intuitive hand motions for users. The implementation and evaluation design of the proposed solution was guided by the formulation of three research questions:
\begin{itemize}
    \item \textbf{RQ1:} Is a template-based gesture matching approach for picking up virtual objects reasonable? 
    \item \textbf{RQ2:} Can users define and use their own gestures without help and technical knowledge?
    \item \textbf{RQ3:} How do custom gestures compare to the state of the art for picking up virtual objects in terms of accuracy, task completion time, and perceived naturalness?
\end{itemize}
A user study was designed and conducted in order to answer each of these questions. The contributions of this paper are as follows:
\begin{itemize}

    \item A comprehensive user study consisting of two experiments that compare three techniques to pick up objects: \begin{itemize}
        \item Controller as a baseline for comparison.
        \item The pinch gesture representing the current state of the art for picking up virtual objects with bare hands
        \item The proposed technique to pick up objects with customised hand gestures.

    \end{itemize}
        \item A system for design, implementation, and prototyping of object tailored hand gestures to pick up virtual objects
\end{itemize}

\section{Background}
\subsection{Virtual Object Manipulation}
Previous work on how to move virtual objects with bare hands in immersive environments has been conducted. \citet{suzuki2014grasping} introduced an AR system to grab a  virtual object with bare hands using the pinch gesture.  The authors generate composite images to achieve occlusion of the virtual object by the real hand. \citet{boonbrahm2014assembly} used the pinch gesture as well in an AR system for assembling small virtual models. Furthermore, the pinch gesture was used by \citet{sorli2021fine} to compare different hand visualisation techniques. Participants had to grab and place big and small cubes. Another example of using the pinch gesture for grabbing virtual objects is the work of \citet{mu2021virtual}. In their work, two different implementations of the pinch gesture are compared. The work of Nguyen \cite{nguyen2022integrating} uses hand features to detect a grasping gesture to overlay tools such as a hammer or screwdriver over the real hands using a MR HMD. \citet{kang_comparative_2020} investigated how the interaction techniques Gaze and Pinch, Direct Touch and Grab, and their novel technique Worlds-in-Miniature compare to each other. It has been found that all techniques have advantages as well as disadvantages. One of the most important findings of the study conducted by Kang et al. is that users prefer a visual guide to the possible interactions regardless of the interaction method used. Grasping virtual objects using bare hands was investigated by Vosinakis et al. \cite{vosinakis2018evaluation}. Specifically, the authors investigated if different visual feedback techniques such as highlighting an object had an impact on usability. The virtual objects could be grabbed and released by closing and opening the hand. Song et al. \cite{song2012handle} used a handle bar as metaphor to manipulate virtual objects with two hands. Khundam and Vorachart et al. \cite{khundam2021comparative} compared hand tracking and controller in a medical training scenario. Users had to pick up virtual objects and interact with virtual elements by using either real hands or controllers. The authors did not find any significant difference in terms of usability. Masurovsky et al. \cite{mti4040091} compared controller and different pick up techniques with the hand tracking device Leap Motion. An important finding was that controller outperformed the other techniques and that controller was not perceived as more natural than hand gesture grasping. Similar results were proposed by Caggianese et al. \cite{caggianese2018vive} where grabbing objects with a controller was compared to a bare handed technique. \citet{OlinDesigning} proposed a system for cross device collaboration in VR where users could pick up virtual objects with their hands. Another interesting approach was introduced by \citet{handinterfaces} where hand gestures are not used to pick up objects but to imitate an object. For example, instead of picking up a virtual scissor, the hand is shaped as a scissor to cut a paper. Hand gestures in the work of \citet{handinterfaces} are defined in a similar template matching approach than the proposed gesture capture technique. \par 
The main difference with the existing works compared to the presented approach is that the gestures used in the aforementioned works to grasp virtual objects are not tailored to the shape of the objects.

\subsection{Gesture Authoring}
This section briefly presents relevant work on the simple creation of gestures that are not necessarily related to the grasping of virtual objects. \citet{10.1145/3411764.3445766} introduced Gesture Knitter, a system to design and implement hand gestures. Users are able to create their own gestures with a visual declarative script. Fine and gross primitives can be combined to create dynamic hand gestures. A system to create and prototype multi-touch gestures was introduced by \citet{10.1145/2207676.2208693}. Kinect Analysis was proposed by \citet{10.1145/2774225.2774846} where motion recordings captured from a RGB-D camera can be inspected and annotated. These recordings can then be used as gestures. AnyGesture \cite{app12041888} introduced rapid prototyping and testing of static and dynamic one-handed gestures. A similar template-matching approach for the gestures presented in this work is used. \citet{ashbrook2010magic} introduced MAGIC to create and prototype gestures with a three-axis accelerometer. \citet{speicher2018gesturewiz} introduced GestureWiz which uses video input data to record and recognise gestures with a consumer-grade webcam. GestuRING from \citet{vatavu2021gesturing} introduced a web based tool to create hand gestures with a finger-worn device. Gestures from a database could be mapped to certain actions. For example, the gestures could be used to navigate a menu when a ring was rotated. Another web based tool is introduced by \citet{magrofuoco2019gestman} for the creation of stroke gestures using a 3D touchpad. A tool to evaluate and create micro gestures was introduced by \citet{li2021design}. The systems presented in this section allow for easy creation of new gestures, but none are designed to make virtual objects naturally tangible by allowing one to create custom hand gestures.

\section{Implementation}
This section describes how the gesture authoring process was implemented. First, the semi-automatic steps for authoring a gesture are described. Next, the architecture of the proposed solution is presented.
\subsection{The Two-Step Gesture Authoring Process}
\begin{figure*}[t]
\centering
\includegraphics[width=1.0\textwidth]{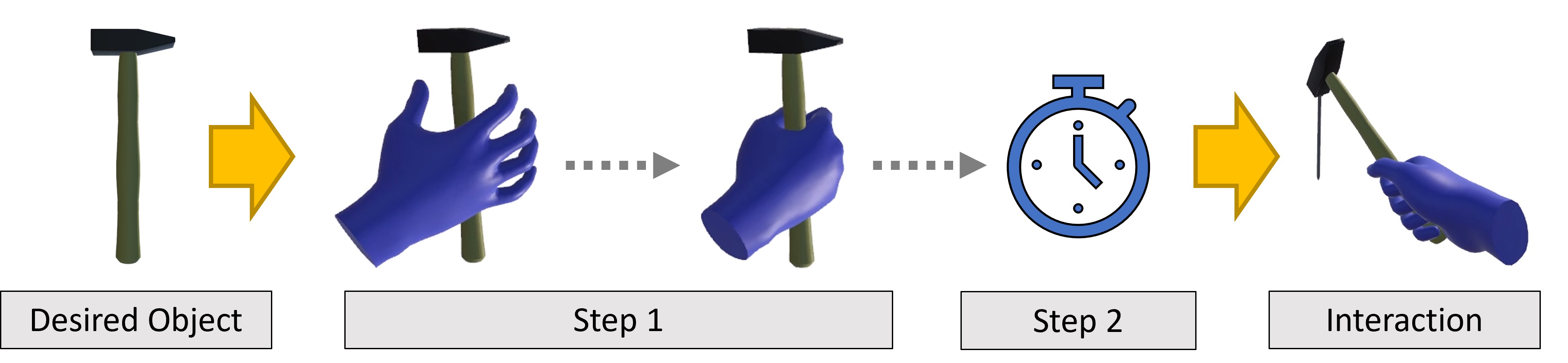}
\caption{The gesture authoring workflow. A chosen 3D object is placed inside the virtual environment. Then, the user has to imitate the grab of the object by shaping the hand as desired. Finally, the hand needs to be in position for three seconds in order to capture the desired grab gesture. In the end, the object can be grabbed and used for more interaction. }
\label{Chapter:GestureAuthoring:fig:workflow}
\end{figure*}

The Gesture Authoring Space uses a simple two-step mechanism in order to allow custom tailored hand object interaction.
\begin{enumerate}
    \item The user needs to place the desired hand near a virtual object, imitating a grab interaction.
    \item After the hand is kept still for three seconds, the gesture is captured and coupled to the object.
\end{enumerate}
This process is depicted in Figure \ref{Chapter:GestureAuthoring:fig:workflow}. Two necessary conditions were identified in order to enable this two-step mechanism: First, the user has to place the hand near an object. The distance between the virtual object and the hand is computed which is then used to recognise if the hand is close enough to the desired object. Secondly, the user needs to keep the position of the hand for a certain amount of time. The user gets visual feedback how long the hand must be maintained in form of a percentage going up from 0\% to 100\%. Once 100\% is reached, i.e. the user kept the hand still for a certain amount of time, the object is attached to the hand and can be released by changing the hand shape to a different pose than the captured gesture (e.g. opening the hand). The first condition is required to store the actual hand shape for the grab gesture. The second condition is necessary in order to capture the gesture when the user finished placing the hand around the object. The desired virtual object is placed on a table within the virtual environment. This table provides information about the two conditions so that the user is informed about the current status of the gesture capturing process. Virtual buttons are placed in front of the user which allows to switch the virtual object to a different one as well as resetting the position and all attached gestures on the current object. In the current implementation only one custom gesture can be attached to an object but this can easily extended to multiple gestures for a single object.

\subsection{System Architecture}
The gesture authoring process requires four principal components: A hand tracking provider, gesture capture, gesture recognition, and the object interaction logic. The interaction capabilities such as attaching a virtual object to the hand were implemented by using a built-in framework of the game engine Unity\footnote{https://unity.com/de}. \par 

The proposed gesture capture and recognition technique was developed through customisation of AnyGesture \cite{app12041888}. 
The recording process will store the currently formed hand shape while the recognising component will compare these stored values with the live hand tracking data. By calculating the Euclidean distance between hand joints of the stored and live data, gestures can be recognised. The similarity $S_g$ of a gesture is recognised by comparing each hand joint (25 provided by the hand tracker) to a stored template by calculating the spatial differences of joints. With $j_c$ being the current joint position provided from the tracker and $j_{gt}$ the joint position in the gesture template as shown in equation \ref{Chapter:GestureAuthoring:eq:templatematching}. 
\begin{equation}
    S_g = \sum_{i=1}^{25} \sqrt{(j_c - j_{gt})^2} \text{, for g = 1,2 \dots , N}
    \label{Chapter:GestureAuthoring:eq:templatematching}
\end{equation}

The hand shape recognition can be done strictly or loosely by adjusting a threshold which will mark when a hand gesture should be detected. The capture process is depicted in Algorithm \ref{Chapter:GestureAuthoring:algo1} and the recognition is explained in Algorithm \ref{Chapter:GestureAuthoring:algo2}. A threshold of 5 cm was used in order to recognise a gesture, i.e. the combined euclidean distance of hand joints ($S_g$) in the current hand tracking frame should not exceed 5 cm to the stored template. This is an empirical value that proved suitable in initial pilot tests, as it was neither too strict nor too loose in recognising certain gestures.\par

\begin{algorithm}
	\caption{Capturing Gestures}
	\label{Chapter:GestureAuthoring:algo1}
	\begin{algorithmic}[1]
		\STATE Desired hand shape is formed around desired object by the user
		\STATE Gesture capture event is triggered by holding hand and fingers still (3 seconds in the user study)
		\STATE Extract joint positions from current hand pose
		\STATE Create new gesture object
		\FOR {Each joint on the hand}
		\STATE Transform joint position from world space to local space
		\STATE Adjust joint position with hand scaling factor
		\STATE Store joint position in gesture object
		\ENDFOR
		\STATE Fill new gesture with the stored values and attach it to the virtual object
	\end{algorithmic} 
\end{algorithm} 
\begin{algorithm}
	\caption{recognising Gestures}
	\label{Chapter:GestureAuthoring:algo2}
	\begin{algorithmic}[1]
	\STATE New hand frame arrives
		\IF {Hand is near an object with attached hand gestures}
    		\STATE Get current hand frame $H_{current}$ from hand tracker
    		\FOR {Each registered gesture in the nearby object}
    		\STATE Set $D_{minimum}$ to Float.Maximum value
    		\FOR {Each joint on the hand}
    		\STATE Transform joint position from world space to local space
    		\STATE Adjust joint position with hand scaling factor
    		\STATE Calculate distance $D$ between stored joint position and current position 
    		\IF {$D>Threshold$}
    		\STATE Discard current gesture $G$  (hand shape is not matching)
    		\ENDIF
    		\STATE Add $D$ to $D_{Sum}$
    		\ENDFOR
    		\IF { $D_{Sum} < D_{minimum}$  }
    		\STATE Set $D_{minimum}$ to $D_{Sum}$ (store the smallest distance)
    		\ENDIF
    		\ENDFOR
    		\IF {$G$ is not discarded AND $D_{Sum}$ is the lowest between gestures}
    		\STATE Current gesture is detected and virtual object is grabbed
    		\ENDIF
		\ENDIF
	\end{algorithmic} 
\end{algorithm}

\begin{figure}[htbp]
\centering
\includegraphics[width=0.3\textwidth]{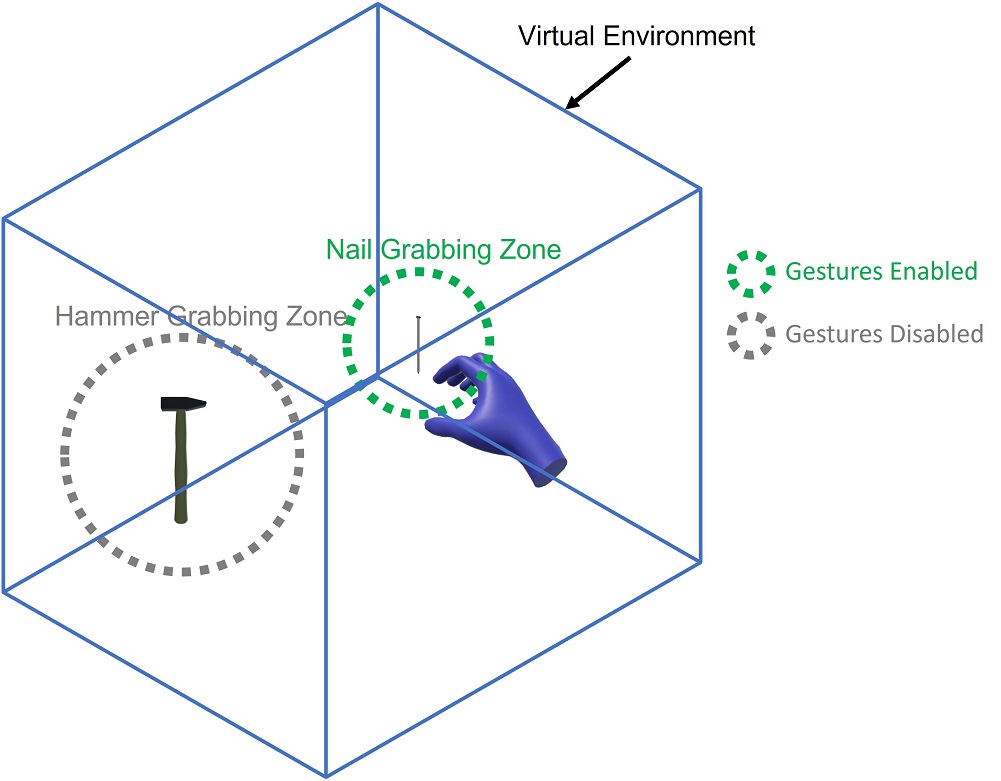}
\caption{Example of context aware gestures: While being near a virtual object, certain hand gestures are enabled. If the hand is too far away from a virtual object, hand gestures for this object are disabled.}
\label{Chapter:GestureAuthoring:fig:GestureZone}
\end{figure}

Hand gestures for grabbing virtual objects in the Gesture Authoring Space are attached to specific objects rather than a global storage for gestures. Virtual objects enter a hover state if a users hand is near them (hover radius is about 10 cm) and gets "unhovered" if the hand is too far away (depicted in Figure \ref{Chapter:GestureAuthoring:fig:GestureZone}). Once a tangible  object is hovered, all attached gestures for grabbing it are registered to the gesture recogniser. The gesture recogniser will then search for hand shapes associated to the gestures in each hand tracking frame that arrives. Unhovering the object will unregister all gestures attached to the tangible object. This allows many different gestures to be recognised without worrying about falsely recognised gestures since the gestures can only be activated when they are actually desired. If two similar gestures are detected, the gesture with the smallest spatial difference will be triggered.
\begin{figure*}[t]
\centering
\includegraphics[width=1.0\textwidth]{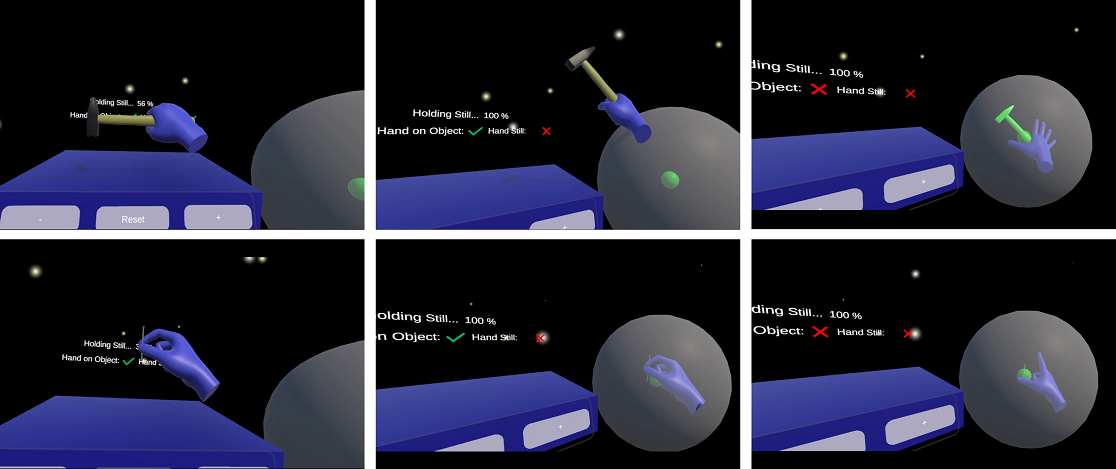}
\caption{The Gesture Authoring Space. A user wraps the hand around a virtual object (left). After three seconds, the object can be grabbed and moved (middle). The object is then be placed inside a transparent sphere (right) for the experimental task to measure accuracy and task completion time. The depicted objects are Hammer (top) and Nail (bottom).}
\label{Chapter:GestureAuthoring:fig:GestureAuthoring}
\end{figure*}

\section{Evaluation}
The Gesture Authoring Space (See Figure \ref{Chapter:GestureAuthoring:fig:GestureAuthoring}) was evaluated with two experiments. The main objective of the evaluation is to compare the created gestures with the pinch gesture. The pinch gesture is considered because it is widely used and the standard solution for many hardware manufacturers regarding bare handed interaction. This includes the Microsoft Hololens 1 and 2, Leap Motion, Meta Quest 2. Controller was added as a control variable as it is the current gold standard for interaction in VR.  When grasping virtual objects, user preference, accuracy and task completion time are the most important metrics found in the literature. Furthermore, the research questions mentioned in section \ref{Chapter:GestureAuthoring:sec:Introduction} should be answered. Therefore, the following evaluation was considered:\par
First, since the Gesture Authoring Space should use template matching as gesture recognition, it was investigated how a template-based gesture compares in relation to the pinch gesture and the controller. Accuracy and task completion time are used as metrics in this evaluation step. The accuracy is determined by how close a virtual object was placed to its target in the following experiments. The task completion time represents the speed a task was considered completed. It should be mentioned that there was no gesture authoring in this step and the used gestures were recorded/created by one of the authors.\par
Secondly, examination of the results of the first study showed that a template-based gesture performed similar to a pinch gesture and did not differ significantly in accuracy and task completion time. This result led to the conclusion that template-based gestures are promising for picking up virtual objects. Furthermore, the gesture authoring process was evaluated in terms of usability to gain insights if it can be reliably used by participants. It should be investigated if users, especially lay people, can create and use custom tailored hand gestures for grabbing virtual objects. \par 
Furthermore, it should be investigated how the custom tailored gestures compare to the pinch gesture and controller in terms of accuracy, task completion time, usability, and naturalness. 

\begin{figure*}[htbp]
\centering
\includegraphics[width=1.0\textwidth]{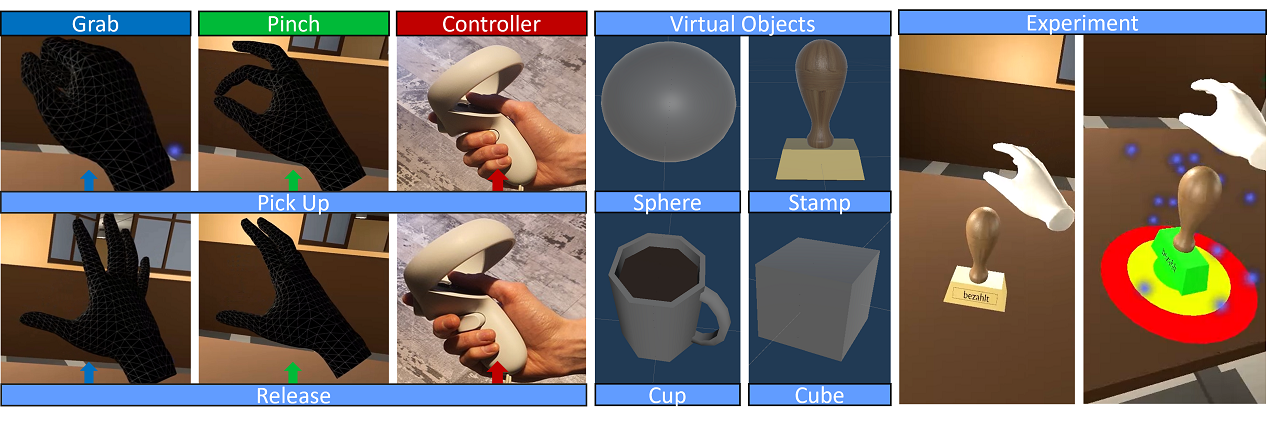}
\caption{The first experiment compares three different pickup techniques for grabbing and placing virtual objects. The gestures Grab and Pinch as well as Controller are compared. Participants are required to place the four virtual objects Sphere, Stamp, Cup, and Cube on the centre of a target. The Grab gesture was implemented using a template matching approach.}
\label{Chapter:GestureAuthoring:fig:Step1_Experiment}
\end{figure*}

\subsection{Apparatus}
The evaluation was performed using a gaming notebook with an Intel Core I7-7820HK, 32 GB DDR4 RAM, Nvidia Geforce GTX 1080 running a 64-bit Windows 10. Hand tracking was realised using the Oculus Quest 2 VR HMD. The game engine Unity was used to develop the system.

\subsection{First Experiment: Compare Template-Based Gestures and Pinch Gesture}

\subsubsection{Implementation of Grasping Techniques}
\label{Chapter:GestureAuthoring:sec:GraspingTechniques}
As grasping virtual objects with controllers or the implementation of a pinch gesture can vary, it is briefly described for reproducibility. 
\par 
\noindent \textbf{Pinch.}
The Pinch gesture was implemented by utilizing thumb and index tip positions provided from the chosen hand tracking solution. The distance between those two points is measured for each frame. A threshold determines if a user is currently pinching. If thumb and index finger are closer than 3 cm to each other, pinching is activated (this is an empirical value). It was found that there was a critical area where noise affected recognition. For example, if the fingers were held about 3 cm apart (2.9 - 3.1 cm), the system would jump between pinch detected and no pinch detected. Therefore, it was introduced that the state of the gesture would only change if the same value was reported for 100 ms which resulted in an overall much smoother user experience. \par 

\noindent \textbf{Grab.}
The Grab gesture was implemented by using the aforementioned gesture capturing process. However, as it is a generic hand gesture not tailored to a specific object, no virtual object was used to capture it. Two different static gestures are required for this approach: One for initiating the grab and one for releasing the object. Therefore, a static gesture resembling a closed hand and a static gesture with a relaxed hand are stored. The gestures are rotation invariant. Detecting a closed hand will initiate a grab event to nearby objects while detecting a relaxed hand will release the currently grabbed object. It was decided to include two static gestures for the release state: One with the hand partially open and one with the hand fully open. Ideally, the gesture with the hand partially open releases the object.\par

\noindent \textbf{Controller.}
Grabbing and releasing an object with the controller is performed by pressing the grip button on the VR controller. This is also depicted in Figure \ref{Chapter:GestureAuthoring:fig:Step1_Experiment}.

\subsubsection{Objectives}
The main objective of this evaluation was to ensure that template-based gestures can reliably be used. Especially lay people who never wore an HMD should be able to perform the given tasks and pick up virtual objects. Additionally, the template-based gesture should perform at least similar to the pinch gesture, which is usually the standard gesture for hand-gesture interfaces. Accuracy and task completion time are used as metrics.
\subsubsection{Users}
A total of 18 users (6 female) participated in the experiment (Age $\mu$ = 33.5). A 5-Point Likert scale (ranging from 0 to 5) was used to assess the users subjective VR experience which resulted in an average of $\mu$ = 2.1 (higher value means more experience).
\subsubsection{Task}
The experiment was conducted as within-subject design. Users sat in front of a virtual desk and had to place a virtual object which appeared in front of them. A target appeared on the table were users had to first grab the virtual object and then place it on the target. Depending on how close the users placed the object, the object gave visual feedback on how close it was to the center. Red for far, yellow for near, and green for being very close to the center of the target. The participants were instructed to place the virtual object as fast and as close as possible to the center of the target. The three interaction techniques Grab, Pinch, and Controller were performed in succession. Users got a quick introduction how to grasp an object with each technique and had time to practice each technique for a short time. Figure \ref{Chapter:GestureAuthoring:fig:Step1_Experiment} depicts this experiment.

\subsubsection{Procedure}
The order of techniques were counterbalanced using the Balanced Latin Square algorithm. Each technique could be practiced for a short time by participants. Four objects had to be placed 10 times respectively, resulting in a total of 120 placed objects for each participant.
\subsubsection{Quantitative Results}
\begin{figure}[htbp]
\centering
\includegraphics[width=0.47\textwidth]{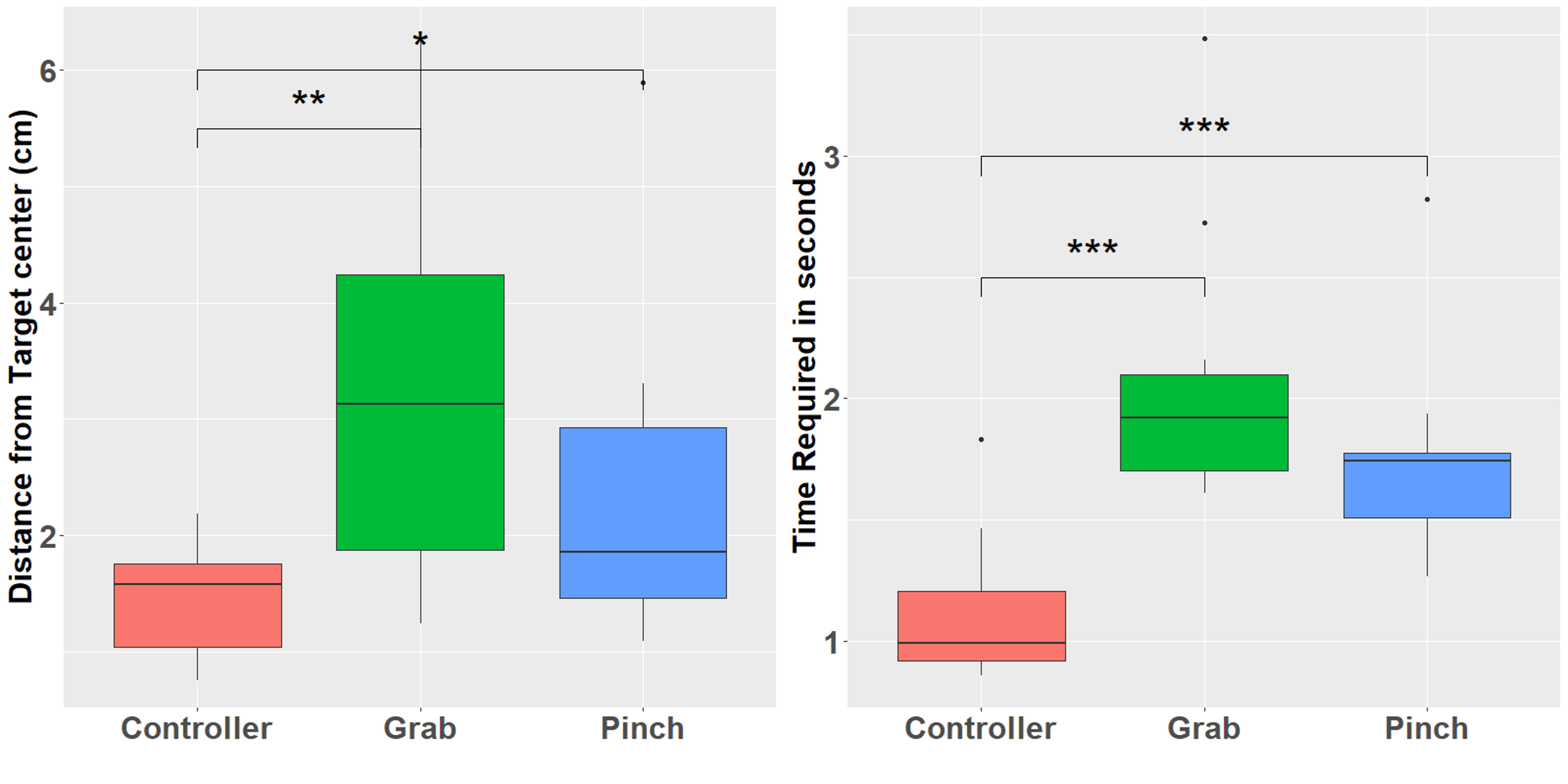}
\caption{First Experiment: Three interaction techniques are compared to each other by grabbing and placing virtual objects. Participants have to place the objects Sphere, Stamp, Cup, and Cube at the centre of a shown target. Accuracy is shown left and the  task completion time on the right. Significance Levels: *** = 0.001; ** = 0.01; * = 0.05.}
\label{Chapter:GestureAuthoring:fig:Dataplot}
\end{figure}

In total, 2.160 virtual objects have been placed in this first experiment of the evaluation (720 per technique). The dependent variables \textit{Accuracy} and \textit{Task Completion Time} are shown in Figure \ref{Chapter:GestureAuthoring:fig:Dataplot}. Levene's test assured the homogeneity of the input data ($p > 0.05$) and therefore one-way ANOVA was used for statistical analysis. The ANOVA result ($F(2,51) = 16.95 ; p < 0.001$) showed significant differences between the techniques in \textit{Task Completion Time}.  \textit{Accuaracy} showed significant differences as well with the ANOVA result ($F(2,51) = 7.108 ; p < 0.01$). Tukey's Honest Significant Difference (TukeyHSD) was used as post hoc analysis of the data. Controller was significantly faster in picking up and subsequently placing an object as compared to the other techniques.  Controller was also significantly more accurate than Grab and Pinch. Grab and Pinch are not significantly faster or more accurate compared to each other.
\subsubsection{Qualitative Results}
\begin{figure}
\centering
\includegraphics[width=0.47\textwidth]{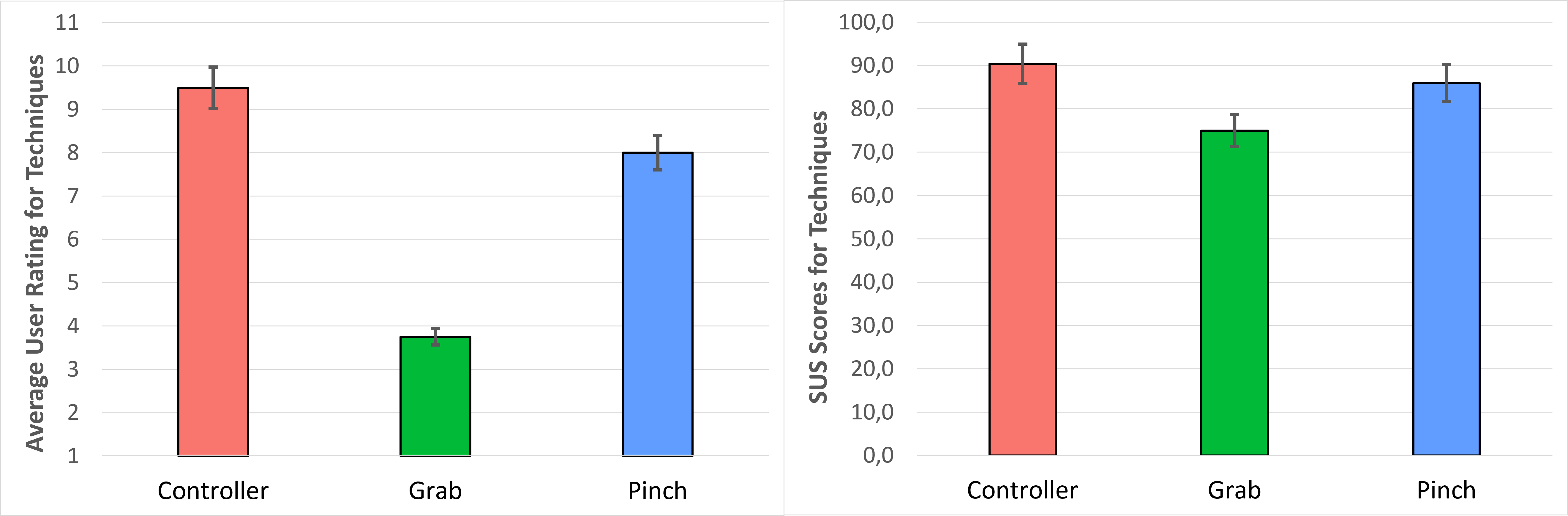}
\caption{First Experiment: Average user rating for each technique on a scale of 1 - 10 (left) and SUS scores (right).}
\label{Chapter:GestureAuthoring:fig:Step1_Qualitative}
\end{figure}

To evaluate the subjective user experience, participants had to answer the System usability scale (SUS) \cite{brooke1996sus,brooke2013sus}. Each time a task was completed with a technique by placing 40 virtual objects, the SUS questionnaire appeared in the virtual reality. After all three techniques were performed, a final questionnaire was shown to the user. This questionnaire allowed users to rate each technique from 1 (bad) to good (10). The SUS scores are the following: Controller 90; Grab 75; Pinch 86. The subjective user rating ranked Grab the worst with an average of 3.75, followed by Pinch with an average of 8 and Controller was rated best by users with an average rating of 9.5. The results are shown in image \ref{Chapter:GestureAuthoring:fig:Step1_Qualitative}.

\subsection{Second Experiment (Part 1): Gesture Authoring}

The second experiment was performed three months after the first and consisted of two parts. Five participants from the first study also participated in this experiment.
\subsubsection{Objectives}
The main objective of the first part of this experiment was to investigate how users can create custom tailored gestures for virtual objects. Specifically, it was investigated if users can create and reuse hand gestures without expert knowledge.
\subsubsection{Users}
A total of 18 users (10 female) with age ranging between 18 and 62 participated in the experiment (Age $\mu$ = 29.7). A 5-point Likert scale (ranging from 0 to 5) was used to assess the users subjective VR experience which resulted in an average of 1.83 (higher value means more experience). Only three participants rated their VR experience higher than two.
\subsubsection{Task}
The experiment was conducted as within-subject design. Users sat in a virtual space with minimal surroundings. A virtual object was placed in front of them. This object was used to start the gesture authoring process. Participants were instructed to wrap their hands around the object as if they would grab them in real life. Then, users had to wait three seconds in order to capture the desired gesture. After the three seconds, the gesture was captured and the object was attached to the users hand. It is to note that these three seconds were empirically determined and could be reduced. If the hand was not kept still, the gesture authoring process is aborted and restarted. To determine if a hand is kept still, the palm and all individual finger tips should not move farther than 1 cm in 10 consecutive hand tracking frames. Changing the hand shape releases the object. Users were allowed to grab and place the object in a transparent sphere which made it disappear. This way, users had a reason to grasp an object and were prepared for the task of grasping and placing in the next part of the experiment. The sphere had a diameter of 50 cm while the largest object (hammer) was about 25 cm long. Three buttons within reach of the user allowed to change the virtual object. Two buttons allowed the user to switch to the next or previous object while the third would reset the current object to its initial position and removes all hand gestures attached to it. A total of eight different objects were used in the experiment: Nail, Cube, Small Cube, Hammer, Ball, Plate, Cylinder, and Paper (as shown in Figure \ref{fig:teaser}). These objects were chosen in order to include big and small objects as well as simple and complex geometric shapes. The gesture authoring task is depicted in Figure \ref{Chapter:GestureAuthoring:fig:GestureAuthoring}.

\subsubsection{Procedure}
Once the user put on the VR HMD, they were immediately in the Gesture Authoring Space. The users could try out different gestures and objects without a strict time limit. However, the experimenter kept this to a maximum of 15 minutes. After the participants captured at least one gesture for each object, a questionnaire was shown to the participants. This questionnaire targeted the subjective user experience.
\subsubsection{Results}
\begin{figure}[htbp]
\centering
\includegraphics[width=0.47\textwidth]{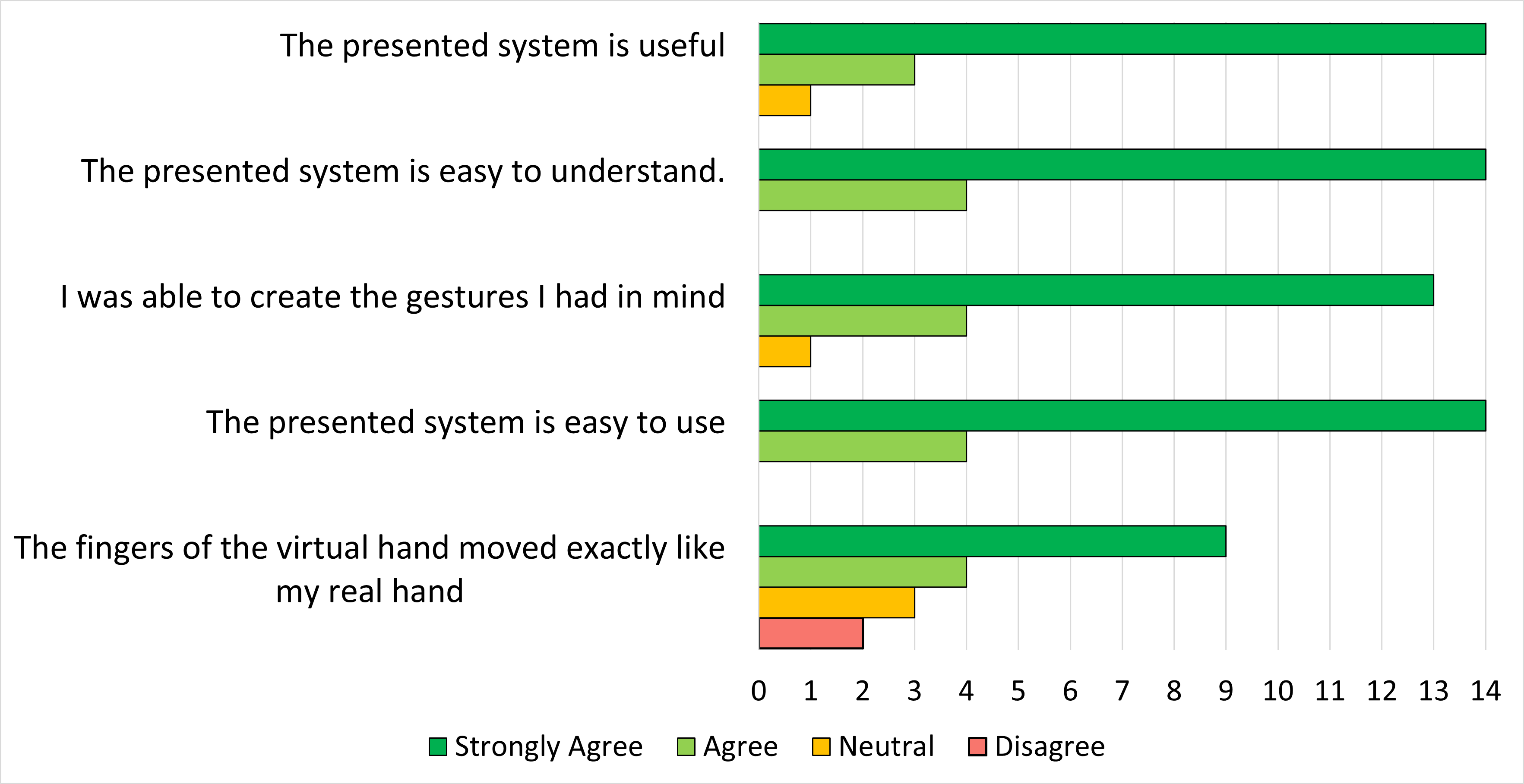}
\caption{Questionnaire results after participants performed the gesture authoring task.}
\label{Chapter:GestureAuthoring:fig:Step2Qualie}
\end{figure}

The following questions were asked after a participant successfully performed the gesture authoring task (5-Point Likert scale ranging from 0 to 5):
\begin{itemize}
    \item \textbf{Q1: }The presented system is useful.
    \item \textbf{Q2: }The presented system is easy to understand.
    \item \textbf{Q3: }I was able to create the gestures I had in mind.
    \item \textbf{Q4: }The presented system is easy to use.
    \item \textbf{Q5: }The fingers of the virtual hand moved exactly like my real hand.
\end{itemize}
In general, the Gesture Authoring Space was well received by users who found it useful, easy to understand and use, and were able to create the gestures they had in mind. The results of the questionnaire are depicted in Figure \ref{Chapter:GestureAuthoring:fig:Step2Qualie}.

\subsection{Second Experiment (Part 2): Evaluate Custom Tailored Gestures}
\subsubsection{Objectives} The main objective of the second part of this experiment was to investigate how custom tailored gestures perform compared to a controller and the pinch gesture. The pinch and controller grasping is implemented as stated in section \ref{Chapter:GestureAuthoring:sec:GraspingTechniques}. The experiment is similar to part one, but the generalised grab gesture was replaced with hand gestures created from the gesture authoring process. The three techniques Controller, Pinch, and Custom Gesture are compared in terms of accuracy, task completion time, usability, and naturalness. Part two of the experiment was conducted directly after the first.

\subsubsection{Task} The experiment was conducted as within-subject design. As in the first part of the experiment, participants sat in a virtual space with minimal surroundings and a virtual object was placed in front of them. This object should be grabbed with one of the three interaction techniques.  Participants had to grab the object in front of them and place it in a transparent sphere, the target area, to measure the accuracy of a technique. The subjects got a quick introduction how to grasp an object with each technique and had time to practice each technique for a short time. The participants were told to move the virtual object as close as possible to the center of the transparent sphere, which was indicated with an opaque point in the middle of it. Once an object is placed inside the target area, it will disappear after one second. The sphere was the same as in experiment part one and had a diameter of 50 cm. The target area was always within reach of the user but changed its position subsequently after placing an object. This was used to check whether the gestures are still reliable when the user extends his/her arm or moves it from left to right and vice versa.  Eight different objects should be grabbed and placed three times with each technique. This resulted in 72 placed objects for each participant (1.296 objects placed in total). The objects are: Nail, Cube, Small Cube, Hammer, Ball, Plate, Cylinder, and Paper (As shown in Figure \ref{fig:teaser}).

\subsubsection{Procedure} The order of techniques was counterbalanced using the Balanced Latin Square algorithm. Participants had to complete the grab and place task for a technique and then the SUS questionnaire was answered within VR. The questionnaire was answered either with the controller or with hand gestures, depending on which technique was previously used. A final questionnaire is shown after the last SUS questionnaire. The aim of this questionnaire is to allow participants to rate each technique on a scale from 1-10 and rate how natural the interaction was on a 5-Point Likert scale (ranging from 0 to 5).

\subsubsection{Quantitative Results}
\begin{figure}[htbp]
\centering
\includegraphics[width=0.47\textwidth]{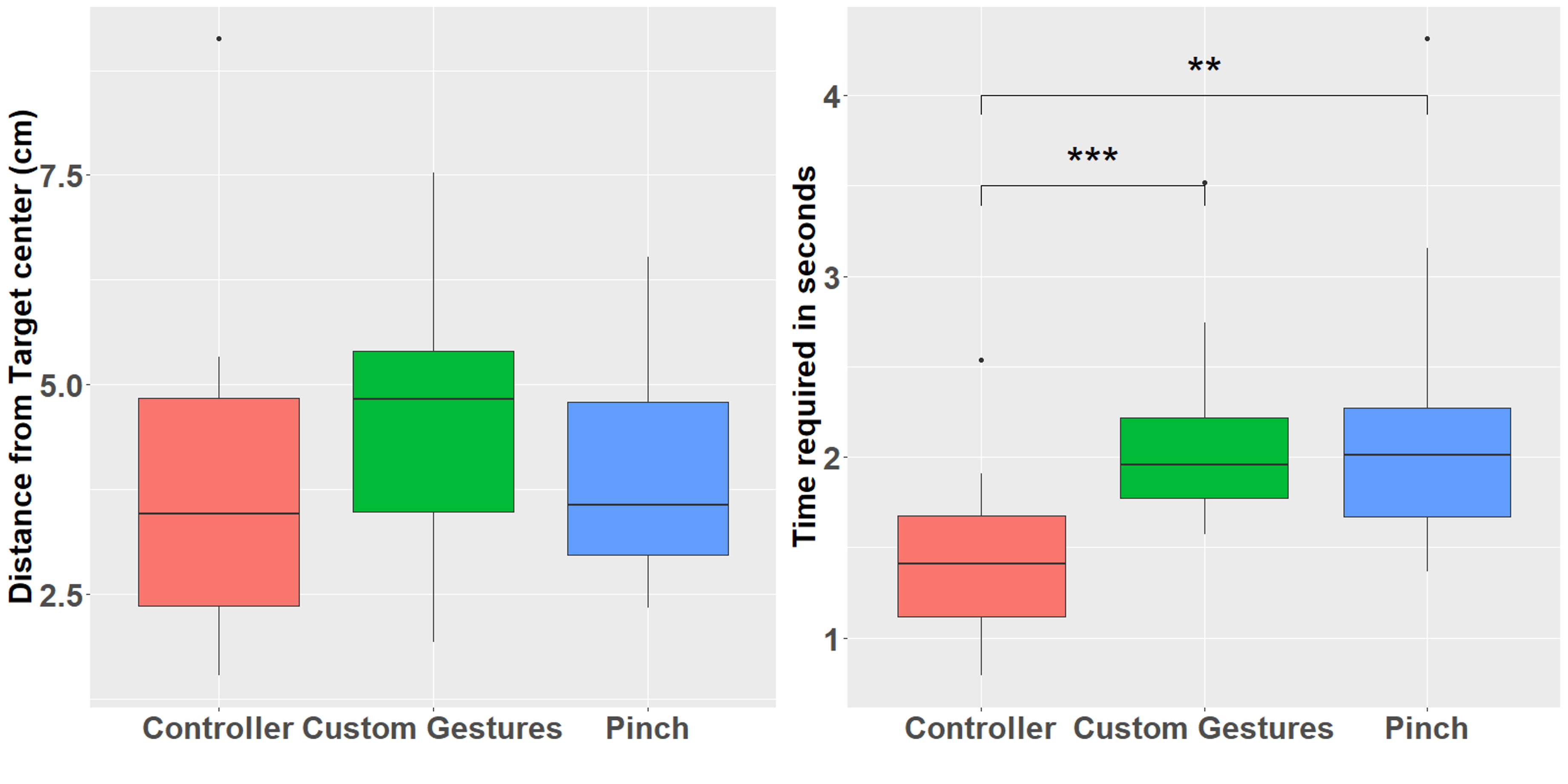}
\caption{Quantitative results of the second experiment. Custom gestures were created by the participants following the gesture authoring process. The accuracy of techniques is depicted on the left and task completion time on the right. Significance Levels: *** = 0.001; ** = 0.01; * = 0.05.}
\label{Chapter:GestureAuthoring:fig:Step3_AccuracyAndTCC}
\end{figure}

Once a participant placed an object, the distance between the center of the object and the target area is stored. Furthermore, the time between grabbing an object and placing it inside the target area is recorded as well.
The dependent variables \textit{Accuracy} and \textit{Task Completion Time} are shown in Figure \ref{Chapter:GestureAuthoring:fig:Step3_AccuracyAndTCC}. Levene's test assured the homogeneity of the input data ($p > 0.05$) and therefore one-way ANOVA was used for statistical analysis. Tukey's Honest Significant Difference (TukeyHSD) was used as post hoc analysis of the data. 
\par Regarding \textit{Task Completion Time}, the ANOVA result ($F(2,51) = 8.726 ; p < 0.001$) showed significant differences between the techniques. Controller was significantly faster than Pinch ($p < 0.05$) and Custom Gesture ($p < 0.01$). However, the pinch gesture showed no statistically significant difference towards the custom gestures. In terms of \textit{Accuracy}, the ANOVA result showed no significance between techniques ($p > 0.05$).
Significance between techniques is depicted in Figure \ref{Chapter:GestureAuthoring:fig:Step3_AccuracyAndTCC}. It is to note that the accidental dropping of an object was also recorded. However, not a single object was accidentally dropped by users. 

\subsubsection{Qualitative Results}
\begin{figure}[htbp]
\centering
\includegraphics[width=0.47\textwidth]{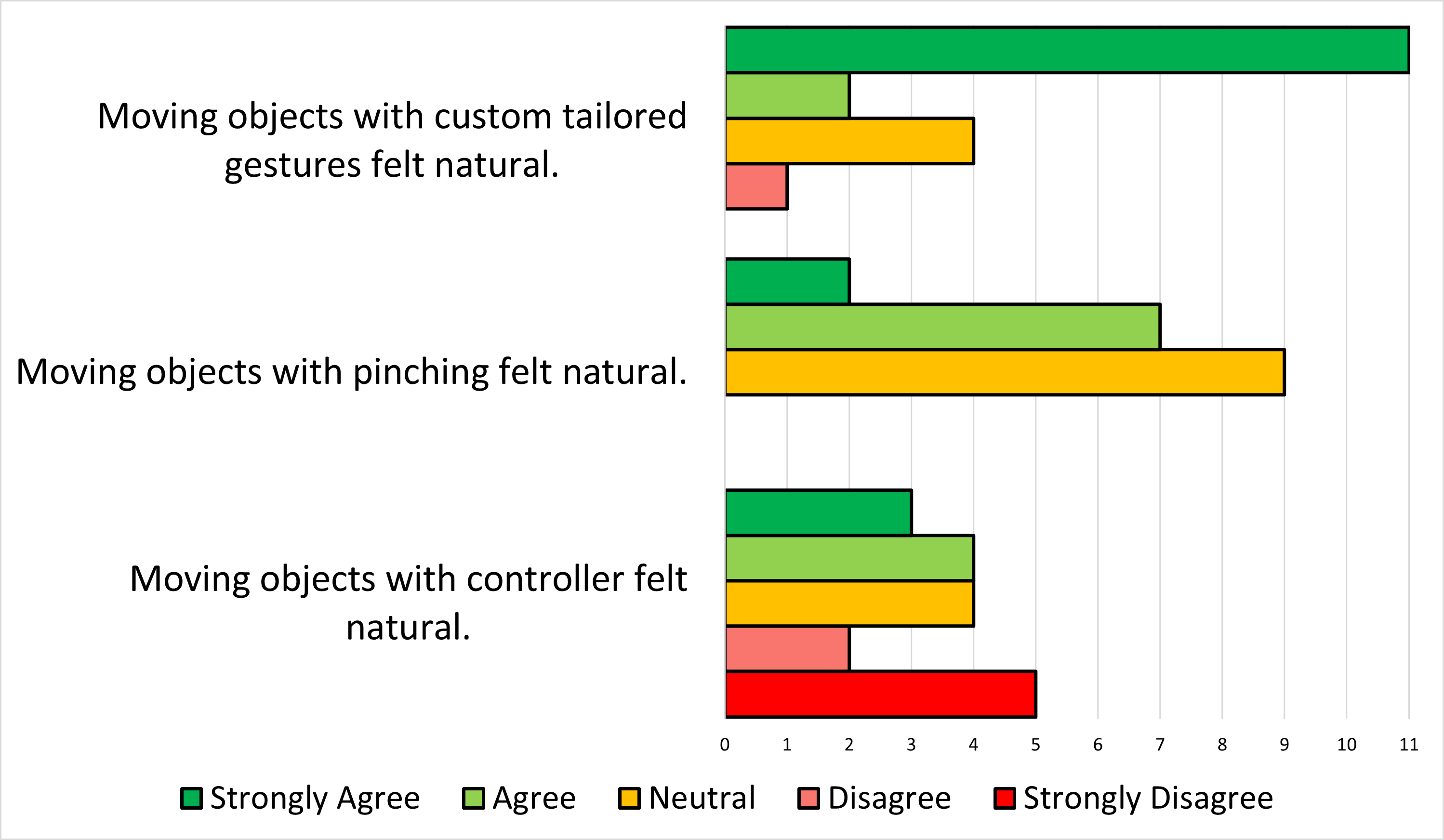}
\caption{Evaluation results of the second experiment. Participants were asked which technique felt natural when grabbing objects.}
\label{Chapter:GestureAuthoring:fig:Step3_Natural}
\end{figure}

Each technique was rated with the SUS by participants. The SUS scores for each technique are the following: Controller 93.3; Custom Gesture 91.1; Pinch 88.5. After performing each technique, a final questionnaire is shown. This questionnaire allowed participants to rate each technique on a scale from 1 - 10. The ratings of techniques are the following: Controller 9.4; Custom Gesture 8.5;  Pinch 8.1.  In addition, for each technique, participants were asked a question about how natural a technique felt on a 5-point Likert scale (ranging from 0 to 5). The participants' answers for perceived naturalness of each technique is depicted in Figure \ref{Chapter:GestureAuthoring:fig:Step3_Natural}. Generally, most users found that moving objects with custom tailored hand gestures felt very natural while using a controller felt unnatural to many. Using the pinch gesture, most users had a neutral attitude.
\begin{figure}
\centering
\includegraphics[width=0.47\textwidth]{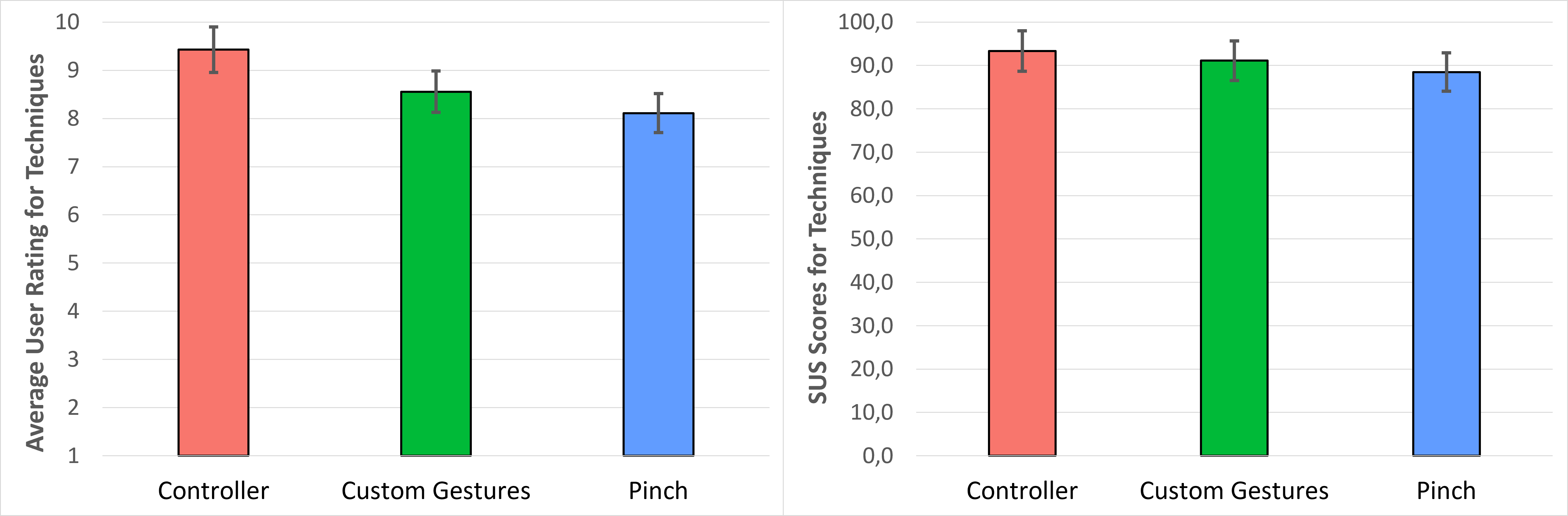}
\caption{Average user rating on a scale of 1 - 10 (left) and SUS scores (right) for the techniques.}
\label{Chapter:GestureAuthoring:fig:Step3_Qualitative}
\end{figure}

\section{Discussion}

The results from the first experiment gave interesting insights into how template based gestures perform compared to controller and the pinch gesture. Not surprisingly, controller was outperforming the bare handed techniques in terms of task completion time and usability. While the quantitative results show that  Grab and Pinch are not significantly different in terms of accuracy and task completion time, the user rating was in favor of Pinch. This rating is likely due to the fact that the pinch gesture is easier to perform and recognise. A template-based gesture requires a relatively precise hand configuration, whereas the pinch gesture depends only on the position of the thumb and index finger. The aim of the experiment was achieved by showing that Pinch and Grab are quantitatively almost indistinguishable, but subjective opinion clearly goes towards the pinch gesture. Therefore, the next experiment investigated how object tailored custom gestures can be designed and implemented to compare it with Pinch and Controller. In particular, it should be investigated whether custom gestures perform better than Grab on subjective assessments when compared to Pinch and Controller.. \par

In the second experiment, the Gesture Authoring Space was well accepted by the participants. Participants created complex gestures which could be recognised without issues. The lowest rating (neutral score) came from a participant who had problems in replicating a gesture for a small object. This was due to the fact that the gesture capturing process was triggered when the virtual hand was not properly mapped to the real hand due to hand tracking failures. The participant then tried to grab the object as desired but the gesture was captured differently.\par
Participants rated the system as useful, easy to understand, and easy to use. Additionally, the participants were able to create the gestures they had in mind. It should be noted that some participants had problems creating complex hand configurations when the view of subsequent fingers was blocked. The gesture authoring process is therefore highly dependent on the quality of the hand tracking system used. This was anticipated before the experiment, which is why question 5 was included in the questionnaire. \par
The second part of this experiment used gestures from the Gesture Authoring Space. The qualitative analysis showed promising results in the custom questionnaires as well as the SUS. Sauro~\cite{sauro2011practical} concludes that a total SUS score above 68 is considered above average. Albert and Tullis~\cite{albert2013measuring} state that a score above 70 can be interpreted as acceptable usability. It was observed that all techniques are above that threshold. Furthermore, the custom tailored hand gestures scored higher in the SUS than the pinch gesture. It should be noted that the SUS was initially designed for more complex system rather than evaluating a simple system for grabbing objects. However, direct comparison of SUS scores between different techniques gives insights how they compare to each other in terms of perceived usability. The subjective scoring of a system on a scale from 1 - 10 was also in favor of the custom tailored gestures as compared to Pinch. The single template-based grab gesture in the first experiment scored relatively low with a score of 3.75. However, the custom tailored gestures scored 8.5 which indicates a significant improvement. Unsurprisingly, controller outperformed the other techniques in terms of accuracy, task completion time, and usability which is consistent with existing research findings \cite{schafer2022comparing,mti4040091,caggianese2018vive}. \par
No accidental drop of a virtual object was recorded with each technique. This is rather surprising since Masurovsky et al. \cite{mti4040091} reported significant differences between objects dropped by different techniques where a hand gesture interface had a mean of 5 dropped objects (out of 30) across participants. This is probably due to todays improved hand tracking and the presented techniques are quite robust. An accidental drop in the proposed system would be recorded once an object is grabbed and placed outside the target area. Once the virtual object was in the target area it counted as a correct placement. Some participants did unintentionally place an object because the hand tracking failed sometimes when participants were reaching out a full arm length. However, it was counted as correct placement since this happened in the target area. These "accidental drops" are therefore reflected in the accuracy as the objects were placed quite far away from the center of the target area. \par 
Regarding naturalness, the custom tailored gestures received best results. Previous studies reported that grabbing virtual objects with bare hands did not feel more natural compared to a controller \cite{mti4040091,khundam2021comparative}. Moving objects with custom tailored gestures presented in this work has received better rating by users in terms of naturalness.  \par 

\subsection{Answering the Research Questions}
\subsubsection{RQ1: Is a template-based gesture matching approach for picking up virtual objects reasonable?}
To answer this question, the first experiment was conducted. It can be said that a template-based hand gesture was on par to the pinch gesture but inferior to the controller. In terms of accuracy and task completion time there was no significant difference to the pinch gesture, showing that it can be a viable bare handed alternative. In terms of usability the template-based gesture was ranked the lowest but still got above average usability results with a SUS score of 75. However, it became only a score of 3.75 out of 10 from users. All of these results indicate that template-based gestures are useful but it is highly depending on which gesture is actually designed. Therefore, the Gesture Authoring Space aims to give users the opportunity to create their own gestures with a template matching approach.
\subsubsection{RQ2: Can users define and use their own gestures without help and technical knowledge?}
Experiment two was conducted to answer this question. Observing the users as well as analysing the questionnaire shows promising results. Even people with no technical background could use the system seamlessly and without noticeable issues. The questionnaire underlines this assumption. Therefore, it can be said that users can define and use their own gesture without help and technical knowlege with the presented system.
\subsubsection{RQ3: How do custom gestures compare to the state of the art for picking up virtual objects in terms of accuracy, task completion time, and perceived naturalness?}
The current gold standard for moving virtual objects is using a controller. As far as the state of the art in moving virtual objects with bare hands is concerned, the pinch gesture has been placed in this position to answer this research question. Controller was not significantly more accurate than the bare handed techniques. However, controller was significantly faster than both bare handed techniques. The custom gestures were not significantly faster or more accurate than the pinch gesture. Users' average rating (from 1 to 10) gave custom gestures a higher rating than Pinch. Additionally, the SUS score of custom gestures was marginally higher than Pinch. However, custom gestures are perceived as more natural than both Pinch and Controller. Therefore, it can be said that the customised gestures from the Gesture Authoring Space are a more natural input compared to the others.
\section{Limitations}
Controller still received best scores in the SUS, accuracy, and task completion time across experiments. Therefore it is still recommended to use controller for simple tasks which require fast movement and high precision. It has yet to be investigated how bare hand interaction compares to a controller in more complex rather than a simple grab and place task.

It should also be noted that experiment one and two had grab and place tasks but differed slightly. The first difference is that the first experiment used a different virtual environment. The second difference is that in the first experiment there were only four objects while in the second experiment there were eight. It was found that for custom tailored gestures there should be more objects in order to properly evaluate custom tailored gestures. This led to different results regarding accuracy, task completion time, and usability for Pinch and Controller in experiment one and two. 
There is also the limitation regarding the number of participants. Due to the COVID-19 pandemic, only a limited amount of participants could be recruited. The Gesture Authoring Space is also limited to one-handed gestures for grasping objects. Grabbing an object with custom tailored gestures for both hands simultaneously is not yet supported.
\section{Conclusion}
This work presents the Gesture Authoring Space to create custom tailored hand gestures for grabbing virtual objects. The proposed solution does not require users to have expert knowledge to design and create gestures they have in mind. A two-step mechanism is used to capture desired hand gestures: 1) Users wrap the hand around this object as if they would grab it in the real world. After the hand is kept still for three seconds the hand gesture is captured. The system was evaluated with two experiments. The first experiment aimed to investigate how the proposed template matching approach compares to state of the art techniques. It was found that a template-based gesture is a viable choice regarding accuracy and task completion time but still lacks behind in usability. Based on these results, experiment two used the template matching approach to record and recognise more natural gestures. The first part of the second experiment evaluated if users can define and use their own gestures without help and technical knowledge. It was found that users with different background knowledge can effectively use the proposed technique and could create the gestures they had in mind. In the second part of this experiment, the custom tailored hand gestures are compared to a controller and the pinch gesture. Gestures created by the Gesture Authoring Space performed better than the initial grab gesture from the first experiment. Specifically, it was found out that controller is still outperforming the bare handed techniques in terms of accuracy, task completion time, and perceived usability. However, the custom tailored gestures outperforms the other techniques regarding naturalness.
\begin{acks}
Part of this work was funded by the Bundesministerium für Bildung und Forschung (BMBF) in the context of ODPfalz under Grant 03IHS075B. This work was also supported by the EU Research and Innovation programme Horizon 2020 (project INFINITY) under the grant agreement ID: 883293.
\end{acks}
\bibliographystyle{ACM-Reference-Format}
\bibliography{sample-base}

%
%
%

\end{document}